\documentclass[12pt,a4paper,oneside,reqno]{amsart}

\usepackage{amssymb,amsmath}
\usepackage{graphicx,psfrag}
\usepackage{stmaryrd} 

\newcommand{\comment}[1]{}

\newcommand{\softcoll}{\text{\sl inecoll}}

\newcommand{\Ob}{\mathrm{IOb}}
\newcommand{\Ph}{\mathrm{Ph}}

\newcommand{\la}{\langle}

\newcommand{\ra}{\rangle}

\newcommand{\pp}{p}  
\newcommand{\qq}{q}  
\newcommand{\vv}{{\vec{v}}}

\newcommand{\Q}{\mathrm{Q}}
\newcommand{\B}{\mathrm{B}}
\newcommand{\W}{\mathrm{W}}
\newcommand{\M}{\mathrm{M}}
\newcommand{\Ib}{\mathrm{Ib}}
\newcommand{\IOb}{\mathrm{IOb}}

\newcommand{\leteq}{\,\mbox{$:=$}\,}

\renewcommand{\and}{\;\land\;}
\newcommand{\setclose}{\}}
\newcommand{\Setclose}{\,\right\}}
\newcommand{\setopen}{\{}
\newcommand{\Setopen}{\left\{\,}

\usepackage{color}
\definecolor{thmcolor}{rgb}{0,0,.4}
\definecolor{remarkcolor}{rgb}{0,.2,0}
\definecolor{proofcolor}{rgb}{.4,0,0}
\definecolor{quecolor}{rgb}{.2,.2,0}
\definecolor{axcolor}{rgb}{.23,0,.23}
\definecolor{axbgcolor}{rgb}{1,.6,1}
\definecolor{defbgcolor}{rgb}{0.9,0.8,0.1}
\definecolor{thmbgcolor}{rgb}{0.8,0.8,1}
\definecolor{rmbgcolor}{rgb}{0.7,1,0.7}
\definecolor{proofbgcolor}{rgb}{1,0.7,0.7}
\definecolor{mg}{rgb}{.6,0,.6}
\newcommand{\mg}{\textcolor{mg}}

\newcommand{\ax}[1]{\textcolor{axcolor}{\ensuremath{\mathsf{#1}}}}
\newcommand{\Ax}[1]{\textcolor{axcolor}{\colorbox{axbgcolor}{\ensuremath{\mathsf{#1}}}}}
\newcommand{\df}[1]{{\bf #1}}

\newcommand{\Dtf}[1]{\setlength{\fboxsep}{2pt}\colorbox{defbgcolor}{#1}\setlength{\fboxsep}{3pt}}
\newcommand{\Df}[1]{\setlength{\fboxsep}{2pt}\colorbox{defbgcolor}
{\ensuremath{#1}}\setlength{\fboxsep}{3pt}}
\newcommand{\Dff}[1]{\setlength{\fboxsep}{1pt}\colorbox{defbgcolor}
{\ensuremath{#1}}\setlength{\fboxsep}{3pt}}

\theoremstyle{definition} \newtheorem{thm}{\colorbox{thmbgcolor}{\textcolor{thmcolor}{Theorem}}}[section]
\theoremstyle{definition} \newtheorem{cor}[thm]{\colorbox{thmbgcolor}{\textcolor{thmcolor}{Corollary}}}
\theoremstyle{definition} 
\theoremstyle{definition} \newtheorem{prop}[thm]{\colorbox{thmbgcolor}{\textcolor{thmcolor}{Proposition}}}

\theoremstyle{remark} \newtheorem{conv}[thm]{\colorbox{rmbgcolor}{\sc\textcolor{remarkcolor}{Convention}}}
\theoremstyle{remark} 
\theoremstyle{definition} 
\theoremstyle{definition} \newtheorem{rem}[thm]{\colorbox{rmbgcolor}{\textcolor{remarkcolor}{Remark}}}

\begin{document}

\title[Relativistic dynamics without conservation postulates]
{Axiomatizing relativistic dynamics without conservation postulates}
\author{H.\ Andr\'eka, J.\ X.\ Madar\'asz, I.\ N\'emeti and G.\ Sz\'ekely}
\thanks{Research supported by
        Hungarian National Foundation for Scientific Research grants
        No T43242, T73601 as well as by Bolyai Grant for Judit X.\ Madar\'asz.}
\date{\today}

\begin{abstract}
A part of relativistic dynamics is axiomatized by
simple and purely geometrical axioms formulated within first-order
logic. A geometrical proof of the formula connecting relativistic
and rest masses of bodies is presented, leading up to a geometric
explanation of Einstein's famous $E=mc^2$. The connection of our
geometrical axioms and the usual axioms on the conservation of mass,
momentum and four-momentum is also investigated.
\end{abstract}

\maketitle

\section{Introduction}

The idea of elaborating the foundational analysis of the logical
structure of spacetime theory and relativity theories (foundation of
relativity) in a spirit analogous with the rather successful
foundation of mathematics was initiated by several authors including
David Hilbert~\cite{Hi02},  cf.\ also \cite[6th problem]{Hi00},
Patrick Suppes~\cite{Sup59}, Alfred Tarski~\cite{HST59} and leading
contemporary logician Harvey Friedman~\cite{FriFOM1},
\cite{FriFOM2}.

There are several reasons for seeking an axiomatic foundation of a
physical theory \cite{suppes}. One is that the theory may be better
understood by providing a basis of explicit postulates for the
theory. Another reason is that if we have an axiom system we can ask
ourselves which axioms are responsible for which theorems. For more
on this kind of foundational thinking called reverse mathematics,
see, e.g., Friedman~\cite{FriFOM1} and Simpson~\cite{S}.
Furthermore, if we have an axiom system for special or general
relativity, we can ask what happens with the theory if we change one
or more of the axioms. That could lead us to a new physically
interesting theory. That is what happened with Euclid's axiom system
for geometry when Bolyai and Lobachevsky altered the axiom of
parallelism which lead to the discovery of hyperbolic geometry.

In the above spirit, in earlier works the Relativity and logic group
of R\'enyi Mathematical Institute in Budapest built up relativity
theories (both special and general) purely in the framework of
first-order logic (\Dtf{FOL}). This foundation of relativity is
elaborated in strict parallel to the success story of the foundation
of mathematics, cf., e.g., \cite{logst}, \cite{anwu}.

Why do we insist on staying within FOL as a framework? For good
reasons, the foundation of mathematics has been carried through
strictly within the framework of FOL. One of these
reasons is that staying within FOL helps us to avoid tacit
assumptions. Another reason is that FOL has a complete inference
system while higher-order logic cannot have one by G\"odel's
incompleteness theorem, see, e.g., \cite[p.505]{vaananen}. For more
motivation for staying within FOL as opposed to higher-order logic,
see, e.g., \cite{AMNsamp}, \cite[Appendix 1: Why exactly
FOL]{pezsgo}, \cite{Ax}, \cite{FeBSL}, \cite{Pam}, \cite{wolenski}.
The same reasons motivate the effort of keeping the foundation of
spacetime and relativity theory within FOL.

In our earlier works we concentrated on the kinematics of relativity
theories. The present paper is devoted to a part of relativistic
dynamics or mechanics. In particular, we present an axiom system
\ax{SpecRelDyn} for relativistic inertial mass. It is an extension
of our earlier axiom system \ax{SpecRel} used for the kinematics of
special relativity. Just as we did in \ax{SpecRel},  we try to keep
our axioms as few as possible and at the same time convincing,
transparent and easy to comprehend even for someone not familiar
with the basic concepts of physics. We also try to keep our axioms
visualizable and purely geometrical. Based on \ax{SpecRelDyn}, we
present a purely geometrical proof for the theorem that relates the
relativistic mass of a moving particle to its rest mass. The usual
approach in standard relativity texts goes by assuming as new axioms
the conservation of relativistic mass and conservation of momentum,
cf.\ d'Inverno~\cite[p.43-36]{d'Inverno} and
Rindler~\cite[pp.108-112]{Rin}. These are very strong assumptions
compared to ours, and by our above mentioned proof, these strong
assumptions are not needed for introducing or explaining
relativistic mass. We base our theory on more basic and more
geometrical axioms. Being more basic and geometrical, these axioms
are also more elementary and more self-evident.

In Section 2 we fix the first-order language for dynamics of special
relativity theory. In Section 3 we recall the streamlined FOL axiom
system \ax{SpecRel} used for kinematics of special relativity theory
from our previous works. In Section 4 we extend \ax{SpecRel} to
cover relativistic dynamics leading to Einstein's famous insight
$E=mc^2$. In Section 5 we present a purely geometric axiom that is
equivalent to conservation of mass and momentum. This axiom is also
proved to be equivalent to the conservation of four-momentum. In
Section 6 we sketch some possible future research directions.

\section{A first-order logic frame for relativity theory}

The motivation for our choice of vocabulary (basic concepts) is
summarized as follows. We represent motion as changing spatial
location in time. To do so, we will have reference-frames for
coordinatizing events (sets of bodies) and, for simplicity, we will
associate reference-frames with certain bodies which we will call
{\em observers}. We visualize an observer  as ``sitting'' in the
origin of the space part of its reference-frame, or equivalently,
``living'' on the time-axis of the reference-frame.
 There will be another special kind of bodies which we will call
{\em photons}. For coordinatizing events, we will use an arbitrary
{\em ordered field} in place of the field of  real numbers. Thus the
elements of this field will be the {\em quantities} which we will
use for marking time and space. In the axioms of dynamics we will
use {\em relativistic masses} of bodies as a basic concept.

Allowing arbitrary ordered fields instead of the field of reals
increases the flexibility of our theory and minimizes the amount of
our mathematical presuppositions, see, e.g., Ax~\cite{Ax} for
further motivation in this direction. Similar remarks apply to our
flexibility oriented decisions below, e.g., the one to treat the
dimension of spacetime as a  variable.

Using observers in place of coordinate systems or reference frames
is only a matter of didactic convenience and visualization. There
are many reasons for using observers (or coordinate systems, or
reference-frames) instead of a single observer-independent spacetime
structure.  One of them is that it helps us to weed  unnecessary
axioms from our theories; but we state and emphasize the logical
equivalence between observer-oriented and observer-independent
approaches to relativity theory elaborated in, e.g., \cite[\S
4.5]{Mphd} and \cite{logst}. Motivated by the above, we now turn to
fixing the first-order language of our axiom systems.

First we fix a natural number $\Dff{d}\ge 2$ for the dimension of spacetime. Our
language contains the following non-logical symbols:
\begin{itemize}
\item unary relation symbols \Dff{\B} (for \df{bodies}), \Dff{\Ob}
(for inertial \df{observers}), \Dff{\Ph} (for \df{photons}) and $\Dff{\Q}$
(for \df{quantities}),
\item binary function symbols  \Dff{+}, \Dff{\cdot} and a binary relation symbol
\Dff{\le}  (for the field operations and the
ordering on $\Q$),
\item a $2+d$-ary relation symbol \Dff{\W} (for \df{world-view relation}),
and
\item a $3$-ary relation symbol \Dff{\M} (for \df{mass relation}).
\end{itemize}

We translate $\B(x)$, $\Ob(x)$, $\Ph(x)$ and $\Q(x)$ into natural
language as ``$x$ is a body,'' ``$x$ is an observer,''  ``$x$ is a
photon,'' and ``$x$ is a quantity.'' (A more careful wording would
be ``$x$ is a possible body," ``$x$ is a possible observer," etc.)
The bodies play the role of the ``main characters'' of our spacetime
models and they are ``observed'' (coordinatized using the
quantities) by the observers. This observation is coded by the
world-view relation by translating $\W(x,y,z_1,\ldots, z_d)$ as
``observer $x$ coordinatizes body $y$ at spacetime location $\langle
z_1,\ldots,z_d\rangle$,'' (that is, at space location $\langle
z_2,\ldots,z_d\rangle$ at instant $z_1$). Finally we use the mass
relation to speak about the relativistic masses of bodies according
to observers by translating $\M(x,y,z)$ as ``$z$ is the mass of body
$y$ according to observer $x$.''

$\B(x)$, $\Ob(x)$, $\Ph(x)$, $\Q(x)$, $\W(x,y,z_1,\ldots, z_d)$,
$\M(x,y,z)$, $x=y$ and $x\leq y$ are the {atomic formulas} of our
first-order language, where $x$, $y$, $z_1$, $\dots$, $z_d$ can be
arbitrary variables or terms built up from variables by using the
field-operations. The \df{formulas} of our first-order language are
built up from these atomic formulas by using the logical connectives
{\em not} (\Dff{\lnot}), {\em and} (\Dff{\land}), {\em or}
(\Dff{\lor}), {\em implies} (\Dff{\Longrightarrow}), {\em
if-and-only-if} (\Dff{\Longleftrightarrow}) and the quantifiers {\em
exists} $x$ (\Dff{\exists x}) and {\em for all $x$} (\Dff{\forall
x}) for every variable $x$.

The \df{models} of this language are  of the form
\begin{equation*}
\langle U; \B, \Ob, \Ph, \Q,+,\cdot,\leq,\W,\M\rangle,
\end{equation*}
where $U$ is a non-empty set and $\B$, $\Ob$, $\Ph$ and $\Q$ are unary
relations on $U$, etc.
A unary relation on $U$ is just a subset of $U$.
Thus we   use $\B$, $\Ob$ etc.\ as sets as well, e.g., we  write
\Dff{k\in \Ob} in place of $\Ob(k)$.

We use the notation $\Dff{\Q^n}\leteq\Q\times\ldots\times \Q$
($n$-times) for the set of all $n$-tuples of elements of $\Q$. If
$\pp\in \Q^n$, then we  assume that $\Dff{\pp}=\langle
p_1,\ldots,p_n\rangle$, that is, ~$p_i\in\Q$ denotes the $i$-th
component of the $n$-tuple $\pp$. We write $\W(m,b,\pp)$ in place of
$\W(m,b,p_1,\dots,p_d)$,  and we  write $\forall \pp$ in place of
$\forall p_1,\dots,p_d$ etc.

We present each axiom at two levels. First we give an intuitive
formulation, then we give a precise formalization using our logical
notation (which can easily be translated into first-order formulas
by inserting the definitions into the formalizations). We seek to
formulate easily understandable axioms in FOL.

The first axiom expresses our very basic assumptions,  such as: both
photons and observers are bodies, etc.

\begin{description}
\item[\Ax{AxFrame}] $\Ob\cup \Ph\subseteq \B$,
$\W\subseteq \Ob \times
\B\times \Q^d$, $\M:\Ob\times \B\rightarrow \Q$ is a function,
$\M(k,b)> 0$ for every observer $k$ and body $b$,
$\B\cap \Q=\emptyset$,  $+$ and $\cdot$ are binary operations and $\le$
is a binary relation on $\Q$.
\end{description}
To be able to add, multiply and compare measurements of observers,
we put an algebraic structure on the set of quantities  by the
next axiom.
\begin{description}
\item[\Ax{AxEOF}]
The \df{quantity part} $\left< \Q; +,\cdot, \le \right>$ is a
Euclidean\footnotemark\ ordered field. \footnotetext{That is, a
linearly ordered field in which positive elements have square
roots.}
\end{description}
For the FOL definition of linearly ordered field, see,
e.g., \cite{Chang-Keisler}. We use the usual field operations \Dff{0,
1, -, /, \sqrt{\phantom{i}}} definable within FOL. We also use the
vector-space structure of $\Q^n$, that is, if $\pp,\qq\in \Q^n$ and
$\lambda\in \Q$, then $\Dff{\pp+\qq, -\pp, \lambda\cdot \pp}\in
\Q^n$; and $\Dff{O}\,\leteq\langle 0,\ldots,0\rangle$ denotes the
\df{origin}. The \df{Euclidean length} of $\pp\in \Q^n$ is defined
as
$\Dff{|\pp|}\leteq\sqrt{\resizebox{!}{8pt}{$p_1^2+\ldots+p_n^2$}}$, for any $n\ge 1$.

\begin{conv}
We treat \ax{AxFrame} and  \ax{AxEOF} as a part of our logical frame
throughout this paper. Hence, without any further mentioning, they
will be always assumed and will be part of every axiom system we
propose herein.
\end{conv}

\section{Kinematics}

In this section we recall the streamlined axiom system \ax{SpecRel}
for kinematics of special relativity theory from our previous works.
We note that \ax{SpecRel} is extended in our works, e.g.,
\cite{Twp}, \cite{logst}, to deal with accelerated observers and
general relativity.

$\Q^d$ is called the \df{coordinate system} and its elements are referred
to as \df{coordinate points}.
We use the notations
\[
\Dff{\pp_\sigma}\leteq\langle p_2,\ldots, p_d\rangle
\text{ and } \Dff{p_\tau}\leteq p_1
\]
for  the \df{space component} and for the \df{time component} of $\pp\in\Q^d$,
respectively.

The \df{event} $ev_k(p)$ is the set of bodies observed by observer $k$  at
coordinate point $\pp$ is, that is,
\[
\Dff{ev_k(\pp)}\leteq\Setopen b\in \B \::\: \W(k,b,\pp)\Setclose.
\]
The \df{world-line} of body $b$ according to observer $k$ is defined as
the set of coordinate points where $b$ was observed by $k$, that is,
\[
\Df{wl_k(b)} \leteq  \Setopen \pp\in \Q^d \: :\: b \in
ev_m(\pp)\Setclose.
\]

Now we formulate our first axiom on observers. (Historically this
natural axiom goes back to Galileo Galilei or even to d'Oresme of
around 1350, but probably it is much more ancient than that, see, e.g., \cite[p.23, \S 5]{AMNsamp}.)

\begin{description}
\item[\Ax{AxSelf}] Each  observer $k$ is motionless in the origin of
the space part of his coordinate system, that is,  his world-line is the
time-axis:
\[
\forall k\in\IOb\quad
wl_k(k)= \setopen\la \lambda,0,\ldots,0\ra\: :\: \lambda\in\Q\setclose.
\]
\end{description}
As a formula of first-order logic this axiom is:
\begin{equation*}
\forall k\in \Ob\ \forall \pp\in\Q^d\quad [\W(k,k,\pp)\
\Longleftrightarrow\
p_2=\ldots=p_d=0].
\end{equation*}

Now we formulate our axiom about the constancy of the speed of
photons. For convenience, we choose  $1$ for this speed.

\begin{description}
\item[\Ax{AxPh}] The world-lines of photons are of slope 1, and
moreover, for every observer, there is a photon through two
coordinate points if their slope is 1:
\begin{multline*}
\forall k\in \Ob\enskip \forall \pp,\qq\in \Q^d\quad [\, |\pp_\sigma-\qq_\sigma|=|\pp_\tau-q_\tau| \iff \\
\qquad \quad\exists ph\in\Ph\quad ph\in ev_k(\pp)\cap ev_k(\qq)
\,].
\end{multline*}
\end{description}
This axiom is a well-known assumption of special relativity, see,
e.g., \cite{logst},  \cite[\S 2.6]{d'Inverno}. In a more careful
interpretation of our logical formalism, instead of ``photons" and
``bodies" we could speak about ``possible world-lines of photons"
and ``possible world-lines of bodies," etc. We chose the present
usage for brevity.

\begin{description}
\item[\Ax{AxEv}] All observers coordinatize the same events:
\begin{equation*}
\forall k,h\in \IOb\enskip \forall \pp\in \Q^d\;\exists \qq\in \Q^d
\quad ev_{k}(\pp)=ev_{h}(\qq).
\end{equation*}
\end{description}

The {\bf world-view transformation}
between the world-views of observers $k$ and $h$  is the set of pairs of
coordinate points $\langle \pp,\qq\rangle$
such that $k$ and $h$ observe the same
event in $p$ and $q$, respectively:
\[
\Df{w^k_h} \leteq \setopen \la \pp,\qq\ra \in\Q^d\times \Q^d\: : \:
ev_k(\pp)=ev_h(\qq)\setclose.
\]

As usual, $\ell$ is called a \df{line} if{}f there are $\pp,\qq\in
\Q^d$ such that $\qq\ne{O}$ and $\ell=\setopen \pp+\lambda \qq\:
:\: \lambda\in \Q\setclose$.

\begin{rem}
\label{specrem}
Assume $d\geq 3$ and \ax{AxSelf}, \ax{AxPh} and \ax{AxEv}.
Then 
\begin{itemize}
\item[(i)] World-view transformations take lines to lines,
see \cite[Thm.11.11.(ii)]{logst}.
\item[(ii)] World-lines of observers are lines by (i) and
\ax{AxSelf}.
\item[(iii)] No observer can travel faster than light, see
\cite[Thm.11.7]{logst}.
\end{itemize}
\end{rem}

By the next axiom we assume that observers use the same units of measurements.
\begin{description}
\item[\Ax{AxSimDist}] Any two observers agree as for the spatial distance
between two events if these two events are
simultaneous  for both of them:
\begin{multline*}
\forall k,h\in\Ob\enskip\forall \pp,\qq,p',q'\in\Q^d\quad
\big[\,\big(ev_k(\pp)=ev_h({p'})\, \land\,
 ev_k(\qq)=ev_h({q'})  \\
\land\,  p_{\tau}=q_{\tau}\, \land\, p'_{\tau}=q'_{\tau}\big)
\  \Longrightarrow\  |\pp_{\sigma}-\qq_{\sigma}|=
|{p'}_{\sigma}-{q'}_{\sigma}|\,\big]. 
\end{multline*}
\end{description}

Let us introduce  an axiom system for special relativistic kinematics:
\begin{equation*}
\boxed{\ax{SpecRel}\leteq\setopen \ax{AxSelf}, \ax{AxPh}, \ax{AxEv},\ax{AxSimDist} \setclose}
\end{equation*}
Let $p,q\in \Q^d$. Then
\begin{equation}
\Df{\mu(p)}:=\left\{
\begin{array}{rl}
\sqrt{p_\tau^2-|p_\sigma|^2 }  & \text{ if }  p_\tau^2-|p_\sigma|^2\ge0 , \\
-\sqrt{|p_\sigma|^2-p_\tau^2 } & \text{ otherwise }
\end{array}
\right.
\end{equation}
is the (signed) {\bf Minkowski length} of $p$ and the {\bf
Minkowski distance} between  $p$ and $q$ is defined as follows:
\begin{equation}
\Df{\mu(p,q)}:=\mu(p-q).
\end{equation}
Function $f:\Q^d\rightarrow \Q^d$ is said to be a {\bf Poincar\'e
transformation} if it is a bijection and it preserves the Minkowski
distance, that is, $\mu\big(f(p),f(q)\big)=\mu(p,q)$ for all $p,q\in
\Q^d$. We note that every Poincar\'e transformation is a linear
transformation composed by a translation. For proof of the following
theorem see Thm.11.10 in \cite{logst}.

\begin{thm}
\label{thmPoi}
\label{spec-thm}
Assume  $d\geq 3$ and \ax{SpecRel}. Then
$w^k_h$ is a Poincar\'e transformation for every $k,h\in\Ob$.
\end{thm}

\noindent
Thus from \ax{SpecRel} if $d\geq 3$, we can deduce the most frequently quoted
predictions of special relativity:
\begin{itemize}
\item[(i)]
``moving clocks slow down,'' \item[(ii)] ``moving meter-rods
shrink'' and
\item[(iii)]
``moving pairs of clocks get out of synchronism.''
\end{itemize}
Moreover, \ax{SpecRel} implies the exact amount of time-dilation,
length-contraction and delay of clocks. So if $d\geq 3$,
\ax{SpecRel} captures the kinematics of special relativity well. For
more detail, see, e.g., \cite{AMNsamp}, \cite{pezsgo}, \cite{logst}.

We often add axioms to \ax{SpecRel} which do not change the
spacetime structure, but are useful auxiliary or bookkeeping axioms.
For example, \ax{AxThEx} below states that each observer can make
thought experiments in which he assumes the existence of ``slowly
moving" observers (see, e.g., \cite[p.622 and Thm.2.9(iii)]{logst}):

\begin{description}
\item[\Ax{AxThEx}] For each observer, in each spacetime location,
in each direction, with any speed less than that of light it is
possible to ``send out" an observer whose time flows ``forwards":
\begin{multline*}
\forall k\in\Ob\enskip\forall \pp,\qq\in\Q^d\enskip \exists
h\in\Ob\enskip 
\big[\, |(\pp-\qq)_{\sigma}|<(\pp-\qq)_{\tau} \ \Longrightarrow\ \\
\pp,\qq\in wl_k(h)\ \mbox{and}\ w^k_h(\qq)_\tau<w^k_h(\pp)_\tau
\,\big]. 
\end{multline*}
\end{description}

\section{Dynamics}
In this section we shall formulate our axioms on dynamics.  The idea
is that we use inelastic collisions for observing (or measuring)
relativistic inertial mass. We could say that relativistic inertial
mass is the quantity that shows the magnitude of the influence of
the body on the state of motion of the body it collides with. The
more a body changes the motion of bodies it collides with, the
bigger its relativistic mass is.

\begin{figure}[ht]
\small
\begin{center}
\psfrag*{q}[l][l]{$q$} \psfrag*{b}[t][t]{$b$} \psfrag*{c}[t][t]{$c$}
\psfrag*{d}[b][b]{$d$} \psfrag*{qt}[r][r]{$q_\tau$}
\psfrag*{out}[b][b]{$out_k(q)$} \psfrag*{in}[t][t]{$in_k(q)$}
\psfrag*{soft}[l][l]{} \psfrag*{k}[l][l]{$k$}
\includegraphics[keepaspectratio, width=\textwidth]{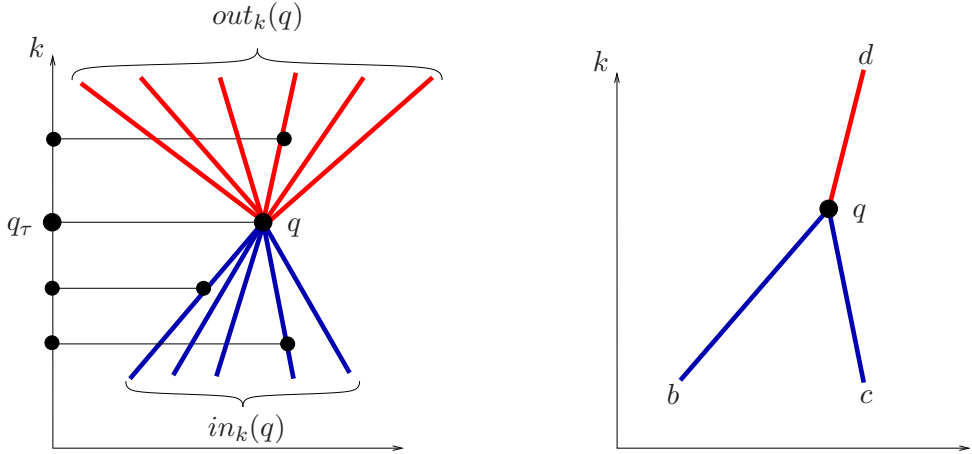}
\caption{\label{inecoll} Illustration of relations $in_k(q)$, $out_k(q)$ and $inecoll_k(b,c:d)$ }
\end{center}
\end{figure}

To formulate our axioms on relativistic mass, first we define
inelastic collisions. The sets $in_k(q)$ of incoming bodies and
$out_k(q)$ of outgoing bodies of the collision at coordinate point
$q$ according to observer $k$ are defined as bodies whose lifelines
``end" and ``start" at $q$ respectively (see Fig.\ref{inecoll}):
\begin{eqnarray*}
\Dff{in_k(q)}  & \leteq & \setopen b\in\B\: :\: q\in wl_k(b)\wedge
\forall p\in
wl_k(b)\ [p_\tau<q_\tau\vee p=q]\setclose,\\
\Dff{out_k(q)} & \leteq & \setopen b\in\B\: :\: q\in wl_k(b)\wedge
\forall p\in wl_k(b)\ [p_\tau>q_\tau\vee p=q]\setclose.
\end{eqnarray*}
Bodies $b$ and $c$ \df{collide inelastically} originating body $d$
according to observer $k$, in symbols {$\Df{\softcoll_k(b,c:d)}$},
if{}f $b\neq c$ and there is a coordinate point $q$ such that
$in_k(q)=\{b,c\}$ and $out_k(q)=\{d\}$, see the right-hand side of
Fig.\ref{inecoll}.

Recall that by \ax{AxFrame}, $\M:\Ob\times \B\rightarrow\Q$ is a
function and $\M(k,b)>0$ for every observer $k$ and body $b$. If $k$
is an observer and $b$ is a body then we call $\Dff{m_k(b)}\leteq
\M(k,b)$ the \df{relativistic mass} of body $b$ according to
observer $k$, or equivalently, ``\dots in the world-view of $k$".

\begin{figure}[ht]
\bigskip
\small
\begin{center}
\psfrag*{t}[r][r]{$t$} \psfrag*{b}[r][r]{$b$}
\psfrag*{ab}[r][r]{$\forall b$} \psfrag*{ad}[l][l]{$\forall d$}
\psfrag*{c}[l][l]{$c$} \psfrag*{ac}[lb][lb]{$\forall c$}
\psfrag*{k}[r][r]{$k$} \psfrag*{ak}[r][r]{$\forall k$}
\psfrag*{c1}[t][t]{$cen_k(b,c)$} \psfrag*{lb}[rb][rb]{$loc_k(b,t)$}
\psfrag*{lc}[lb][lb]{$loc_k(c,t)$} \psfrag*{mb}[lt][lt]{$m_k(c)$}
\psfrag*{mc}[rt][rt]{$m_k(b)\ \ $}
\psfrag*{mk}[b][b]{$cen_k(b,c,t)$} \psfrag*{m}[t][t]{$cen_k(b,c)$}
\includegraphics[keepaspectratio, width=\textwidth]{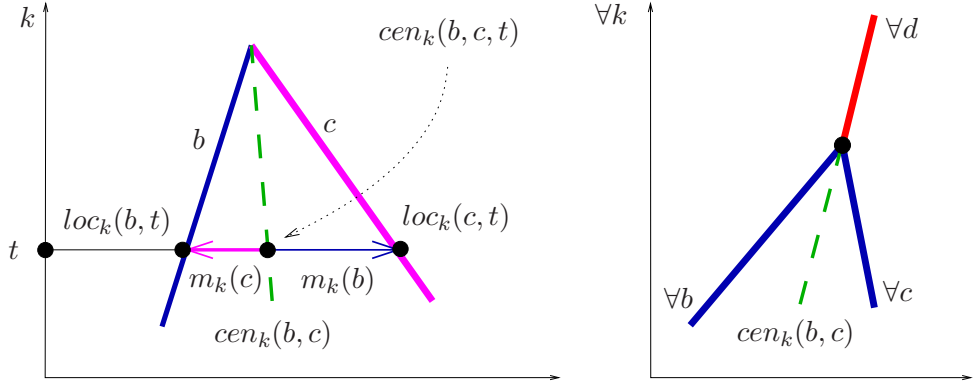}
\caption{\label{med4} Illustration of $cen_k(b,c,t)$, $cen_k(b,c)$
and of axiom \ax{AxCenter}}
\end{center}
\end{figure}

 The \df{spacetime location}  {$\Dff{loc_k(b,t)}$}
of body $b$ at time instance $t\in \Q$ according to
observer $k$ is defined to be
the coordinate point $p$ for which $p\in wl_k(b)$ and $p_\tau=t$ if there is
such a unique $p$, and it is undefined otherwise, see Fig.\ref{med4}.

The \df{center of mass} $\Dff{cen_k(b,c,t)}$ of bodies $b$ and $c$
at time instance $t$ according to observer $k$ is defined to be the
coordinate point  $q$  such that  $q_\tau=t$  and  $q$ is the
point on the line-segment between $loc_k(b,t)$ and $loc_k(c,t)$
whose distances from these two end-points have the same proportion
as that of the relativistic masses of $b$ and $c$; and it is closer
to the ``more massive" body, i.e.:
\begin{equation*}
m_k(b)\cdot\big(loc_k(b,t)-cen_k(b,c,t)\big) =
m_k(c)\cdot\big(cen_k(b,c,t)-loc_k(c,t)\big)
\end{equation*}
if $loc_k(b,t)$ and $loc_k(c,t)$ are defined, and $cen_k(b,c,t)$ is
undefined otherwise,
see Fig.\ref{med4}.
We note that an explicit definition for $cen_k(b,c,t)$ is the following:
\[
cen_k(b,c,t)= \frac{m_k(b)}{m_k(b)+m_k(c)}\cdot loc_k(b,t)+
          \frac{m_k(c)}{m_k(b)+m_k(c)}\cdot loc_k(c,t),
\]
(if $loc_k(b,t)$ and $loc_k(c,t)$ are defined and $cen_k(b,c,t)$ is
undefined otherwise). The \df{center-line of mass}  of bodies $b$
and $c$  according to observer $k$ is defined as
\[
\Df{cen_k(b,c)}:=\setopen cen_k(b,c,t)\: : \: t\in \Q\text{ and
$cen_k(b,c,t)$ is defined}\setclose.
\]
Intuitively, the center-line of mass is the world-line of the center
of mass. The segment determined by $\pp,\qq\in\Q^d$ is defined as:
\[
\Dff{[\pp,\qq\,]}\leteq\setopen
\lambda\cdot\pp+(1-\lambda)\cdot\qq\: : \: \lambda\in\Q,\
0\leq\lambda\leq 1\setclose.
\]
We call $H\subseteq \Q^d$ \df{line segment} if{}f  $H$ is connected
(i.e., $[p,q]\subseteq H$ for all $p,q\in H$), $H$ has at least two
elements, and $H$ is contained in a line.

Bodies whose world-lines are line segments are called \df{inertial
bodies}, and their set is defined as:
\[
\Dff{\Ib}\leteq\setopen b\in \B\: :\: \forall k\in\Ob\quad wl_k(b)
\text{ is a line segment}\setclose.
\]

We note that $cen_k(b,c)$ is a line segment or a point or the empty
set and $wl_k(b)\cap wl_k(c)\subseteq cen_k(b,c)$ for every
$k\in\Ob$ and $b,c\in\Ib$.
\medskip

We are ready now to formalize that the relativistic mass is a
quantity that shows the magnitude of the influence of the body on
the state of motion of the body it collides with.

\begin{description}
\item[\Ax{AxCenter}]
If inertial bodies $b$ and $c$ collide inelastically originating
single inertial body $d$, then the world-line of $d$ is the
continuation of the center-line of mass of $b$ and $c$ (see
Fig.\ref{med4}):
\begin{multline*}
\forall k\in\IOb\enskip \forall b,c,d\in \Ib\quad
[\,\softcoll_k(b,c:d)\ \Longrightarrow\\
 cen_k(b,c)\cup wl_k(d)\subseteq\ell\mbox{ for some line }\ell\,].
\end{multline*}
\end{description}

The main axiom of \ax{SpecRelDyn} is \ax{AxCenter} which, in some
sense, can be taken as the definition of relativistic mass. The
remaining axioms of our axiom system will be simplifying or
book-keeping axioms to make life simpler.

\ax{AxCenter} is an axiom in Newtonian Dynamics, too, where the mass
$m_k(b)$ of a body $b$ is observer-independent in the sense that it
does not depend on the observer $k$. However, in special relativity,
\ax{AxCenter} implies that the mass of a body necessarily depends on
the observer. The reason for this fact is that the simultaneities of
the different observers in special relativity differ from each
other, and this implies that the proportions involved in
\ax{AxCenter} change, too. See Prop.\ref{massdepend} and
Fig.\ref{massdepend-fig} below.
\begin{figure}[ht]
\small
\begin{center}
\psfrag*{k}[r][r]{$k$} \psfrag*{h}[r][r]{\mg{$h$}}
\psfrag*{l'}[lb][lb]{$x_k$} \psfrag*{l}[rb][rb]{\mg{$x_h$}}
\psfrag*{x}[t][t]{$m_k(c)$} \psfrag*{x'}[t][t]{$m_k(b)$}
\psfrag*{y}[l][l]{\mg{$m_h(c)$}} \psfrag*{y'}[b][b]{\mg{$m_h(b)$}}
\psfrag*{b}[r][r]{$b$}\psfrag*{c}[r][r]{$c$}\psfrag*{d}[r][r]{$d$}
\includegraphics[keepaspectratio, width=0.8\textwidth]{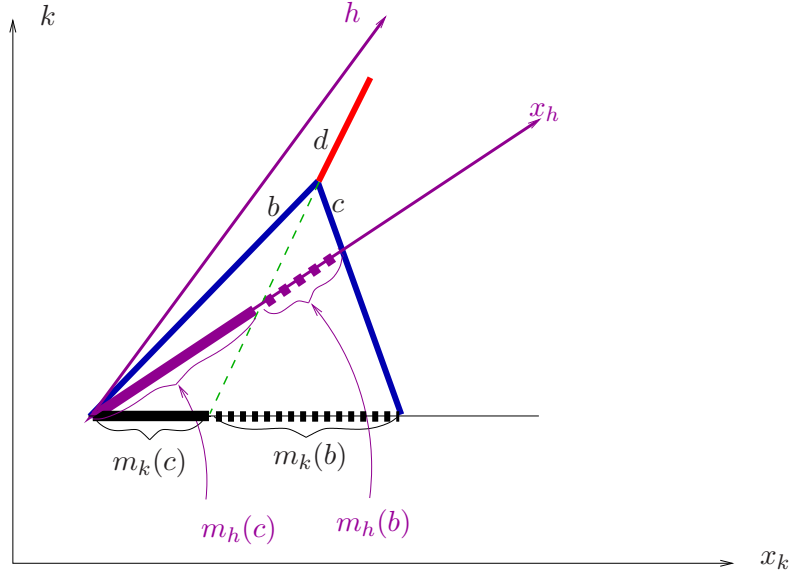}
\caption{\label{massdepend-fig} Illustration for
Prop.\ref{massdepend}. The proportion of the bold and dotted
segments on the horizontal line is different from that on the
slanted one.}
\end{center}
\end{figure}

\begin{prop}
\label{massdepend} Assume \ax{SpecRel} and \ax{AxCenter}. Let
$k,h\in\IOb$, $b,c,d\in\Ib$ be such that $\softcoll_k(b,c:d)$,
 $\softcoll_h(b,c:d)$, $h$ is not at rest w.r.t.\ $k$, and 
 $h$ does not move orthogonally to the collision of $b$ and $c$. Then
\[ \frac{m_k(b)}{m_k(c)}\ne \frac{m_h(b)}{m_h(c)}.
\]
\end{prop}
\noindent We omit the proof of Prop.\ref{massdepend}, but
Fig.\ref{massdepend-fig} is an illustration for it.

The \df{velocity} $\vv_k(b)$ and \df{speed} $v_k(b)$ of body $b$
according to observer $k$ are defined as:
\begin{equation*}
\Df{\vv_k(b)}  \leteq  \frac{\pp_\sigma-\qq_\sigma}{p_\tau-q_\tau},
\text{ for } \pp,\qq\in wl_k(b)\text{ with } p_\tau\neq q_\tau,\text{ and }
\Df{v_k(b)}  \leteq  |\vv_k(b)|
\end{equation*}
if $wl_k(b)$ is a subset of a line and contains coordinate-points
$p$ and $q$ with $p_\tau\neq q_\tau$, and they are undefined
otherwise.

The \df{rest mass} {$\Dff{m_0(b)}$} of body $b$ is defined to be
$\lambda\in \Q$ if (1) there is an observer according to which $b$ is at
rest and the relativistic mass of $b$ is $\lambda$, and (2) for every
observer according to which $b$ is at rest the relativistic mass of
$b$ is $\lambda$, that is, $m_0(b)=\lambda$ iff
\[
\exists k\in\IOb\ (v_k(b)=0\wedge m_k(b)=\lambda)\ \wedge\ \forall
k\in\IOb\ (v_k(b)=0\Longrightarrow m_k(b)=\lambda).
\]

By Rmk.\ref{specrem}, assuming $d\geq 3$, \ax{AxSelf}, \ax{AxEv} and
\ax{AxPh}, if the rest mass of body $b$ is defined,  $b$ is
slower than light, that is, $v_k(b)$ is defined and $v_k(b)<1$ for
every observer $k$. In particular, photons do not have rest masses,
but see Remark~\ref{discussion}.(2) below.

\begin{conv}
We use the equation sign ``='' in the sense of existential equality
(of partial algebra theory \cite{ABuN}), that is, $\alpha=\beta$
abbreviates that both $\alpha$ and $\beta$ are defined and they are
equal. See \cite[Conv.2.3.10, p.31]{Mphd} and \cite[Conv.2.3.10,
p.61]{pezsgo}.
\end{conv}

We have seen that \ax{AxCenter} implies that the relativistic mass
$m_k(b)$ has to depend on both $b$ and $k$. The next axiom states
that the relativistic mass of a body depends at most on its rest
mass and its velocity.

\begin{description}
\item[\Ax{AxSpeed}] The relativistic masses of two inertial bodies
are the same if both of their rest masses and speeds are equal:
\begin{multline*}
 \forall k\in\IOb\ \forall b,c\in\Ib \\
 \big[\big(m_0(b)=m_0(c)\ \wedge\ v_k(b)=v_k(c)\big)\ \Longrightarrow\
m_k(b)=m_k(c)\big].
\end{multline*}
\end{description}

Our last axiom on dynamics states that each observer can make
experiments in which he makes inertial bodies of arbitrary rest
masses and velocities inelastically collide:

\begin{description}
\item[\Ax{Ax\forall inecoll}] For every observer, every kind of possible
inelastic collision is realized by inertial bodies having rest mass:
\begin{multline*}
 \forall k\in\Ob\ \forall v_1,v_2\in\Q^{d-1}\ \forall m_1, m_2\in \Q
\quad  \big(\, |v_1|<1\wedge |v_2|<1 \\
 \wedge m_1>0 \wedge m_2>0 \quad\Longrightarrow\quad
\exists b,c,d\in\Ib\ [\, \softcoll_k(b,c:d)\;\\
  \wedge\; \vv_k(b)=v_1\; \wedge\; \vv_k(c)=v_2\; \wedge\;
m_0(b)=m_1\; \wedge\; m_0(c)=m_2]\,\big).
\end{multline*}
\end{description}

Let us extend \ax{SpecRel} with the axioms of dynamics above.
\begin{equation*}
 \boxed{\ax{SpecRelDyn} \leteq
\setopen\ax{AxCenter}, \ax{AxSpeed},\ax{Ax\forall
inecoll},\ax{AxThEx}\setclose \cup \ax{SpecRel}}
\end{equation*}
We note that \ax{SpecRelDyn} is provably consistent. Moreover it has
non-trivial models, see Prop.\ref{rem}.

The following theorem gives the connection between the rest mass and the
relativistic mass of an inertial body. Its conclusion is a well known result of
special relativity. We will see that our theorem is stronger than the
corresponding result in the literature since it contains fewer assumptions.
\begin{thm}
\label{thm1}
Assume $d\geq 3$ and \ax{SpecRelDyn}.
Let $k$ be an observer and $b$ be an inertial body having rest mass.
Then
\[
m_0(b)={\sqrt{1-v_k(b)^2}}\cdot m_k(b).
\]
\end{thm}

\noindent
\begin{proof}\label{thm1-proof}
Let $k$ be an observer and let $a$ be  an inertial body having rest
mass. Let $v\leteq v_k(a)$, $m_0\leteq m_0(a)$ and $m(v)\leteq
m_k(a)$. We would like to prove that $m_0=\sqrt{1-v^2}\cdot m(v)$.
It holds if $v=0$ by the definition of rest mass. Now assume that
$v\neq 0$. We are in the world-view of observer $k$. Let inertial
bodies $b$ and $c$ collide inelastically originating inertial body
$d$ such that the rest masses of $b$ and $c$ are $m_0$ the speed of
$b$ is $v$ and the speed of $c$ is $0$. See Fig.\ref{dyn1}.
\begin{figure}[ht]
\small
\begin{center}
\psfrag*{ell}[r][r]{$\ell$} \psfrag*{A}[r][r]{$A$}
\psfrag*{B}[lb][lb]{$B$} \psfrag*{C}[rb][rb]{$C$}
\psfrag*{D}[t][t]{$D$} \psfrag*{b}[r][r]{$b$} \psfrag*{c}[l][l]{$c$}
\psfrag*{d}[b][b]{$d$} \psfrag*{A'}[l][l]{$A'$}
\psfrag*{B'}[lb][lb]{$B'$} \psfrag*{C'}[rb][rb]{$C'$}
\psfrag*{D'}[lb][lb]{$D'$} \psfrag*{E'}[r][r]{$E'$}
\psfrag*{mv}[t][t]{$m(v)$} \psfrag*{m0}[t][t]{$m_0$}
\psfrag*{0}[l][l]{$0$} \psfrag*{-1}[l][l]{$-1$}
\psfrag*{v}[r][r]{$-\sqrt{1-v^2}$}
\psfrag*{text1}[t][t]{{coordinate-system of $k$}}
\psfrag*{text2}[t][t]{{coordinate-system of $h$}}
\psfrag*{k}[t][t]{$k$} \psfrag*{l}[t][t]{$k'$}
\includegraphics[keepaspectratio, width=\textwidth]{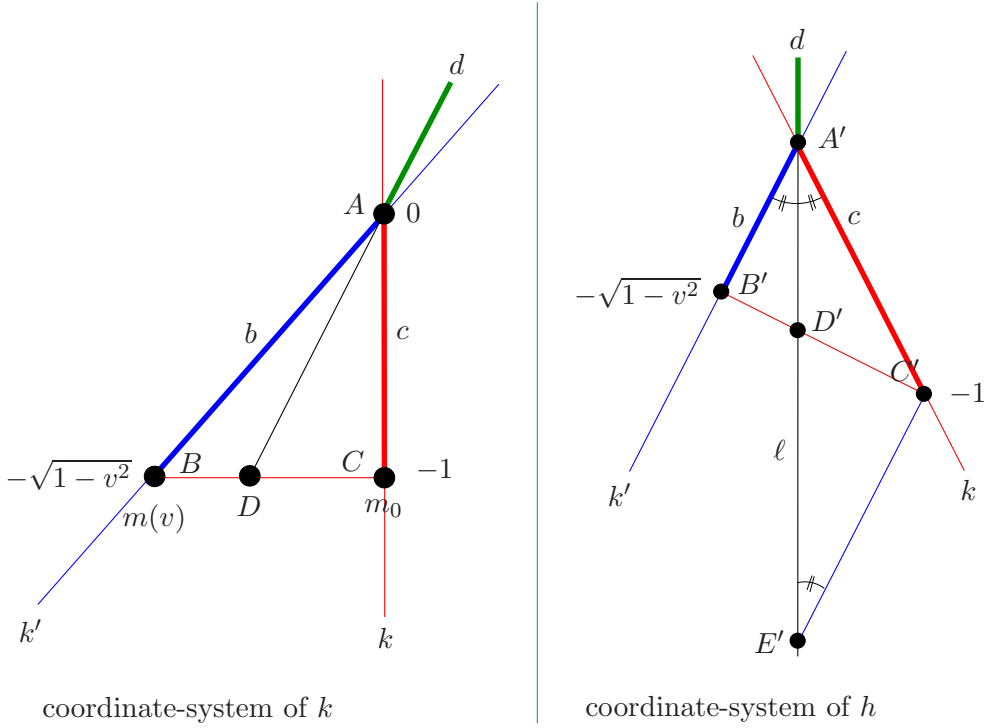}
\caption{\label{dyn1} Illustration for the proof of Thm.\ref{thm1} }
\end{center}
\end{figure}
Such $b$, $c$ and $d$ exist by \ax{Ax\forall inecoll}. There are
distinct points $B$ and $C$ on the world-lines of $b$ and $c$,
respectively, such that $B_\tau=C_\tau$. Let such $B$ and $C$ be
fixed and let $t\leteq B_\tau=C_\tau$. Let $D$ be the center of mass
of $b$ and $c$ at $t$. The relativistic masses of $b$ and $c$
according to $k$ are $m(v)$ and $m_0$, respectively, by \ax{AxSpeed}
and the definition of rest mass. Let $|\pp\qq|:=|\pp-\qq|$. By
definition of center of mass, $m(v)\cdot |BD|=m_0\cdot |DC|$. Thus
\begin{equation}
\label{e1} m_0=\frac{|BD|}{|CD|}\cdot m(v).
\end{equation}
 Let $A$ be the point where the world-lines of $b$, $c$ and $d$ meet.
 By \ax{AxCenter},  $cen_k(b,c)\cup wl_k(d)\subseteq AD$. Let $k'$ be an observer such that
$v_{k'}(b)=0$. Such a $k'$ exists since $b$ has rest mass. We can
assume that the clocks of $k$ and $k'$ show $0$ at $A$, that is,
$A_\tau=w^k_{k'}(A)_\tau=0$, and the clock of $k$ shows $-1$ at $C$,
that is, $C_\tau=-1$. By applying the ``time-dilation theorem'' of
\ax{SpecRel} (see \cite[Thm.11.6.(2)]{logst}) we get that the clock
of $k'$ shows $-\sqrt{1-v^2}$ or $\sqrt{1-v^2}$ at $B$. We can
assume that the clock of $k'$ shows $-\sqrt{1-v^2}$ at $B$.

By \ax{AxThEx} there is an observer $h$ for which $b$ and $c$ have
opposite velocities and $\softcoll_h(b,c:d)$. Let such an $h$ be
fixed.

The world-view transformation $w^k_h$ between the world-views of $k$
and $h$  is an affine transformation, that is, a linear
transformation composed by a translation by
\cite[Thm.11.10.]{logst}. Thus $w^k_h$ takes lines to lines.

Let us turn our attention to the world-view of $h$. See the
right-hand side of Fig.\ref{dyn1}. Let $A'$, $B'$, $C'$ and $D'$ be
the $w^k_h$ images of $A$, $B$, $C$ and $D$, respectively. Since
$w^k_h$ is an affine transformation,
\begin{equation}
\label{a1} \frac{|BD|}{|CD|}=\frac{|B'D'|}{|C'D'|}.
\end{equation}

We will prove that
\begin{equation}
\label{e2} \frac{|B'D'|}{|C'D'|}=\frac{|A'B'|}{|A'C'|}.
\end{equation}

Let $\ell$ be the line parallel to the time-axis $\bar t$ and
passing through $A'$.  Since the rest masses and the speeds of $b$
and $c$ coincide, their relativistic masses coincide by
\ax{AxSpeed}. Therefore $cen_h(b,c)\subseteq \ell$. By
\ax{AxCenter}, $wl_h(d)\subseteq \ell$. The world-view
transformation takes lines to lines and world-lines to world-lines.
Thus $w^k_h$ takes $wl_k(d)\subseteq AD$ to $wl_h(d)\subseteq\ell$.
Therefore $D'$ is the intersection of $\ell$ and $B'C'$.

Let $E'\in A'D'$ be such that $E'C'$ is parallel to $A'B'$. The
triangles $B'D'A'$ and $C'D'E'$ are similar. Thus
\begin{equation}
\label{e3} \frac{|B'D'|}{|C'D'|}=\frac{|A'B'|}{|E'C'|}.
\end{equation}
Since $b$ and $c$ have opposite speeds and $A'B'$ is parallel with
$C'E'$, angles $E'A'C'$ and $A'E'C'$ are congruent. Thus
$|E'C'|=|A'C'|$. By this and (\ref{e3}), we conclude that (\ref{e2})
above holds.

The clocks of $k'$ and $k$ show $0$ at $A'$, the clock of $k'$ shows
$-\sqrt{1-v^2}$ at $B'$ and the clock of $k$ shows $-1$ at $C'$.
The speeds of $k$ and $k'$ coincide for $h$. Thus the clocks of $k$
and $k'$ slow down with the same rate for $h$ by
\cite[Thm.11.6.(2)]{logst}. Therefore
\begin{equation}
\label{a2} \frac{|A'B'|}{|A'C'|}={\sqrt{1-v^2}}.
\end{equation}
By (\ref{e1}), (\ref{a1}), (\ref{e2})  and (\ref{a2}), we get that
\[
m_0={\sqrt{1-v^2}}\cdot m(v);
\]
and that is what we wanted to prove.\end{proof}

\begin{rem}\label{discussion}
(1) The conclusion of Thm.\ref{thm1} fails if we omit any one of the
axioms $\ax{AxCenter}, \ax{AxSpeed},\ax{Ax\forall
inecoll},\ax{AxThEx}$ from \ax{SpecRelDyn}. However, it remains true
if we omit \ax{AxSimDist} and weaken $\ax{Ax\forall inecoll}$ and
\ax{AxThEx} to the following two axioms, respectively:
\begin{description}
\item[\Ax{Ax\exists inecoll}] According to every observer,
for every inertial body $a$ having rest mass there are inertial
bodies $b$ and $c$ colliding inelastically originating an inertial
body such that $a$, $b$ and $c$ have the same rest masses, $a$ and
$b$ have the same speeds and the speed of $c$ is $0$:
\begin{multline*}
 \forall k\in\IOb\ \forall a\in \Ib\ \exists b,c,d\in \Ib\ \
\big(\, m_0(a)=m_0(a)\ \Longrightarrow \\ [ m_0(a)=m_0(b)=m_0(c)\
\wedge
 v_k(b)=v_k(a)\ \wedge\  v_k(c)=0\ \wedge\ \\ \softcoll_k(b,c:d)]\, \big).
\end{multline*}
\item[\Ax{AxMedian}]
For every two inertial bodies colliding
inelastically, there is an observer for which these two inertial
bodies have opposite velocities and collide inelastically:
\begin{multline*}
\forall k\in\Ob\  \forall b,c,d\in \Ib \quad
\big[\softcoll_k(b,c:d)\ \Longrightarrow\\
 \exists h\in\IOb\quad\big(\vv_h(b)=-\vv_h(c)\ \wedge\
\softcoll_h(b,c:d)\big)\big].
\end{multline*}
\end{description}
\smallskip

\noindent (2) According to our definition, photons do not have rest
masses because no observer sees them at rest, by \ax{AxPh}. However,
they do have nonzero relativistic masses, by \ax{AxFrame}. In the
light of Thm.1\ref{thm1} then it is natural to extend the rest mass
concept for photons as  $m_0(ph)=0$  for all  $ph\in \Ph$. This is
often done in the physics literature. One could then say that
``photons could be regarded as pure energy," because they have zero
rest masses.
\end{rem}
\bigskip

\begin{figure}[ht]
\small
\begin{center}
\psfrag*{a}[br][br]{$\forall a$} \psfrag*{b}[t][t]{$b$}
\psfrag*{c}[t][t]{$c$} \psfrag*{eb}[br][br]{$\exists b$}
\psfrag*{ec}[lb][lb]{$\exists c$} \psfrag*{ed}[lt][lt]{$\exists d$}
\psfrag*{ab}[br][br]{$\forall b$} \psfrag*{ac}[t][t]{$\forall c$}
\psfrag*{ad}[lt][lt]{$\forall d$} \psfrag*{k}[r][r]{$\forall k$}
\psfrag*{d}[b][b]{$d$} \psfrag*{h}[r][r]{$\exists h$}
\psfrag*{text}[r][r]{$m_0(a)=m_0(b)=m_0(c)$}
\includegraphics[keepaspectratio, width=\textwidth]{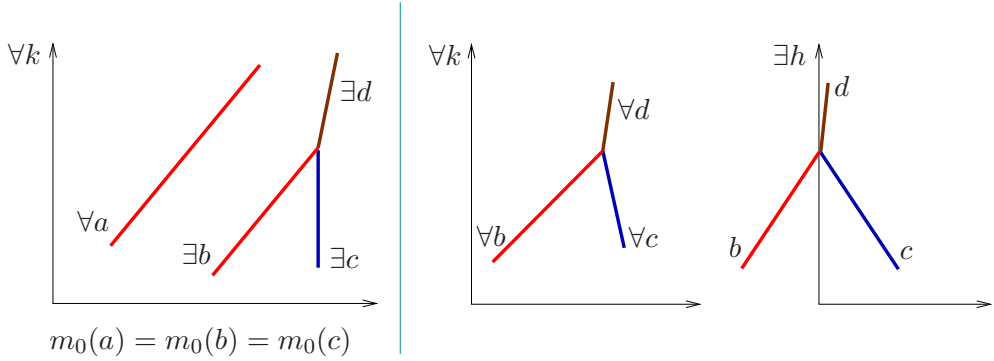}
\caption{\label{pirt} Illustration of axioms \ax{Ax\exists inecoll}
and \ax{AxMedian}}
\end{center}
\end{figure}

\noindent {\em On Einstein's $E=mc^2$}: The conclusion
$m_0(b)={\sqrt{1-v_k(b)^2}}\cdot m_k(b)$ of our Thm.\ref{thm1} above
is used in the relativity textbook Rindler~\cite[pp.111-114]{Rin} to
explain the discovery and meaning of Einstein's famous insight
$\Dff{E=mc^2}$. We could repeat literally this part of the text of
\cite{Rin} to arrive at $E=mc^2$ in the framework of our theory
\ax{SpecRelDyn} based on the axiom \ax{AxCenter}. We postpone this
to section~\ref{cons-sec}, because there we will have developed more
``ammunition," hence the didactics can be made more inspiring.

\section{Conservation of relativistic mass and
linear-momentum}\label{cons-sec}

We can view \ax{AxCenter} as stating that the center of mass of an
isolated system consisting of two bodies moves along a straight line
regardless whether the two bodies collide or not. It is natural to
generalize \ax{AxCenter} to more than two bodies (but permitting
only two-by-two inelastic collisions). Let $\ax{AxCenter_n}$ denote,
temporarily, a version of \ax{AxCenter} which concerns an isolated
system consisting of $n$ bodies. Thus \ax{AxCenter} is just
$\ax{AxCenter_2}$ in this series of stronger and stronger axioms. We
will see that it does not imply \ax{AxCenter_3} (cf.\
Prop.\ref{Prop1}), thus \ax{AxCenter_3} is strictly stronger than
\ax{AxCenter}. However, it can be shown (see \cite{dyn}) that the
rest of the axioms in this series are all equivalent to
$\ax{AxCenter_3}$. This motivates our introducing \ax{SpecRelDyn^+}
by replacing \ax{AxCenter} in \ax{SpecRelDyn} with the stronger
\ax{AxCenter_3}. The theory \ax{SpecRelDyn^+} is still very
geometric and observation-oriented in spirit.

\begin{figure}[ht]
\small
\begin{center}
\psfrag*{a}[br][br]{$\forall a$} \psfrag*{b}[br][br]{$\forall b$}
\psfrag*{c}[bl][bl]{$\forall c$} \psfrag*{d}[bl][bl]{$\forall d$}
\psfrag*{c1}[rb][rb]{$cen_k(a,b,c)$}
\psfrag*{c2}[rb][rb]{$cen_k(a,d)$} \psfrag*{k}[r][r]{$\forall k$}
\includegraphics[keepaspectratio, width=0.7\textwidth]{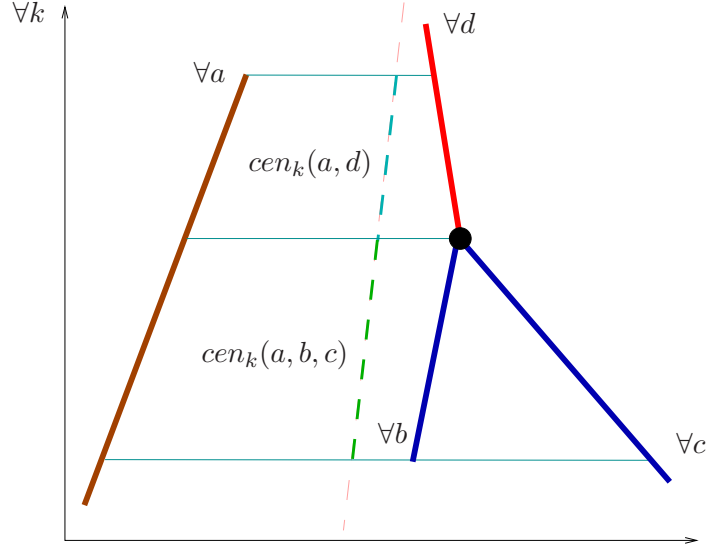}
\caption{\label{centerplus} Illustration of \ax{AxCenter^+}}
\end{center}
\end{figure}

We are going to introduce \ax{AxCenter_3}, we will denote it as
\ax{AxCenter^+}. The center-line of mass \df{$cen_k(a,b,c)$} of
three bodies $a$, $b$ and $c$ according to observer $k$ is defined
in a completely analogous way as for two bodies, as follows. The
\df{center of mass} $\Dff{cen_k(a,b,c,t)}$ of bodies $a,b$ and $c$
according to observer $k$ at time instance $t$ is defined as:
\begin{multline*}
m_k(a)\cdot\big({cen_k(a,b,c,t)}-loc_k(a,t)\big)+
m_k(b)\cdot\big({cen_k(a,b,c,t)}-
loc_k(b,t)\big)\\
 +m_k(c)\cdot\big({cen_k(a,b,c,t)}-loc_k(c,t)\big)=0
\end{multline*}
if $loc_k(a,t)$, $loc_k(b,t)$ and $loc_k(c,t)$ are defined and it is
undefined otherwise. We note that an explicit definition for
$cen_k(a,b,c,t)$ is the following:
\begin{multline*}
 {cen_k(a,b,c,t)}=\frac{m_k(a)}{m_k(a)+m_k(b)+m_k(c)}\cdot loc_k(a,t)+\\
                            \frac{m_k(b)}{m_k(a)+m_k(b)+m_k(c)}\cdot
                        loc_k(b,t)+
                              \frac{m_k(c)}{m_k(a)+m_k(b)+m_k(c)}
\cdot loc_k(c,t).
\end{multline*}
The
\df{center-line of mass}  of bodies $a$, $b$  and $c$  according to
observer $k$ is defined as
\[
\Df{cen_k(a,b,c)}:=\setopen cen_k(a,b,c,t)\: : \: t\in \Q\text{ and
$cen_k(a,b,c,t)$ is defined}\setclose.
\]
\begin{description}
\item[\Ax{AxCenter^+}] If $a$ is an inertial body and inertial
bodies $b$ and $c$ collide inelastically originating inertial body
$d$, then the center-line of $a$ and $d$ is the continuation of the
center-line of $a,b$ and $c$, i.e., there is a line that contains
both the center-line of $a,b$ and $c$ and the center-line of $a$ and
$d$ (see Fig.\ref{centerplus}):
\begin{multline*}
\forall k\in\Ob\ \forall a,b,c,d\in \Ib\quad
 [\softcoll_k(b,c:d)\quad\Longrightarrow\\
 cen_k(a,b,c)\cup cen_k(a,d)\subseteq\ell\mbox{ for some line
}\ell].
\end{multline*}
\end{description}

Let us replace \ax{AxCenter} with \ax{AxCenter^+} in
\ax{SpecRelDyn}:
\begin{equation*}
 \boxed{\ax{SpecRelDyn^+} \leteq
\setopen\ax{AxCenter^+}, \ax{AxSpeed},\ax{Ax\forall
inecoll},\ax{AxThEx}\setclose \cup \ax{SpecRel}}
\end{equation*}
We note that \ax{SpecRelDyn^+} is consistent. Moreover it has
non-trivial models, see Prop.\ref{rem}.
\bigskip

\begin{conv}
Throughout the paper, there appear ``highlighted" statements like
\ax{AxCenter^+} above which associate a name like \ax{AxCenter^+} to
a formula of our first-order language (for \ax{SpecRelDyn}).
It is important to note that these formulas are not automatically
elevated to the rank of an axiom. Instead, they serve as potential
axioms or even as potential statements to appear in theorems, hence
they are nothing but distinguished formulas of our language.
\end{conv}

\ax{AxCenter} determines the velocity of the body emerging from an
inelastic collision, and we will see that \ax{AxCenter^+} determines
also the relativistic mass of the body emerging from the collision.

\begin{description}
\item[\Ax{ConsMass}] Conservation of relativistic mass:
\begin{multline*}
\forall k\in\IOb\ \forall b,c,d\in \Ib\quad  [ \softcoll_k(b,c:d)\
\Longrightarrow\\
 m_k(b)+m_k(c)=m_k(d)].
\end{multline*}
\end{description}

The \df{linear-momentum} of body $b$ according to observer $k$ is
defined to be $m_k(b)\cdot\vv_k(b)$ if $\vv_k(b)$ is defined, and it
is undefined otherwise.

\begin{description}
\item[\Ax{ConsMoment}] Conservation of linear-momentum:
\begin{multline*}
\forall k\in\IOb\ \forall b,c,d\in \Ib\quad  [\softcoll_k(b,c:d)\ \Longrightarrow\\
  m_k(b)\cdot\vv_k(b) +m_k(c)\cdot\vv_k(c) =m_k(d)\cdot\vv_k(d)].
\end{multline*}
\end{description}

The following theorem states that \ax{AxCenter^+} is equivalent to
the conjunction  of \ax{ConsMass} and any of the  two formulas \ax{AxCenter}
and  \ax{ConsMoment}, but it is strictly stronger than any
one of them (Prop.\ref{Prop1}). This means, in some sense, that
\ax{ConsMass} represents the ``difference" between \ax{AxCenter} and
\ax{AxCenter^+}, and the same holds for \ax{ConsMoment}.

\begin{thm}
\label{thm2} Assume \ax{AxSelf}. Items (i)--(iv) below are
equivalent
\begin{itemize}
\item[(i)] $\ax{AxCenter^+}$.
\item[(ii)] $\ax{ConsMass}\wedge\ax{ConsMoment}$.
\item[(iii)] $\ax{ConsMass}\wedge\ax{AxCenter}$.
\end{itemize}
\end{thm}
\noindent The proof of Thm.\ref{thm2} is in \cite{dyn}.

\begin{cor}\label{cor}
Assume \ax{SpecRelDyn^+}. Let $k\in\IOb$, $b,c,d\in \Ib$ and assume
$\softcoll_k(b,c:d)$ and $m_0(b), m_0(c), m_0(d)$ exist. Then
\begin{eqnarray*}
m_k(d)&=&m_k(b)+m_k(c),\qquad  \mbox{but}\\
m_0(d)&>&m_0(b)+m_0(c),\qquad  \mbox{whenever } \vv_k(b)\ne
\vv_k(c)\, .
\end{eqnarray*}
\end{cor}

The proof is in \cite{dyn}, but for the idea of the proof see
below.\bigskip

\noindent {\em Returning to $E=mc^2$}: Cor.\ref{cor} above can be
used for arriving at Einstein's insight $E=mc^2$ analogously to how
it is done in the relativity textbooks Rindler~\cite{Rin} and
d'Inverno~\cite{d'Inverno}. Namely, we have seen above, in
Cor.\ref{cor}, that under appropriate arrangement, rest mass can be
created. Created from what? Well, from kinetic energy (energy of
motion). This points in the direction of Einstein's connecting
mass with energy. In more detail, let us start with two 
bodies $b_1,b_2$ of rest mass $m_0$. Let us accelerate the two
bodies towards each other and let them collide inelastically, so
that they stick together forming the new body ``$b_1+b_2$"
(deliberately sloppy notation). Assume  $b_1+b_2$  is at rest
relative to the observer conducting the experiment. Then the rest
mass $m_0(b_1+b_2)$ is the sum of relativistic masses $m_k(b_1)$ and
$m_k(b_2)$ by Cor.\ref{cor}. Assuming that at collision the speed of
both $b_1$ and $b_2$ was $v$, we have
$m_0(b_1+b_2)=m_0(b_1)/\sqrt{1-v^2}+m_0(b_2)/\sqrt{1-v^2}$, by
Thm.\ref{thm1}, which is definitely bigger than $m_0(b_1)+m_0(b_2)$
if $v\ne 0$. So, rest mass was created from the kinetic energy
supplied to our test bodies $b_1,b_2$ when we accelerated them
towards each other. So far, we have a qualitative argument (based on
our \ax{SpecRelDyn^+}) in the direction that energy (in our example
kinetic) can be ``transformed" to ``create" mass. A quantitatively
(and physically) more detailed analysis of $E=mc^2$ in terms of
Thm.\ref{thm1} is given in \cite[pp.111-114]{Rin} to where we refer
the reader for more detail and for the ``second part" of the
argument. The ``first part" was provided by Thm.\ref{thm1} and
Cor.\ref{cor}.

\medskip
Let $\varphi$ be a formula and $\Sigma$ be a set of formulas.
$\Sigma \models \varphi$ denotes that $\varphi$ is true in all
models of $\Sigma$ (i.e., $\varphi$ is a logical consequence of
$\Sigma$). $\Sigma\not\models\varphi$ denotes that there is a model
of $\Sigma$ in which $\varphi$ is not true.

\begin{prop}\label{Prop1}
\begin{eqnarray*}
\ax{SpecRelDyn} &\not\models&\ax{ConsMass} ,\qquad \text{and} \\
\ax{SpecRelDyn} &\not\models&\ax{ConsMoment}.
\end{eqnarray*}
\end{prop}
\noindent The proof of Prop.\ref{Prop1} is in \cite{dyn}.

In the literature, the conservation of relativistic mass and that of
linear-momentum are used to derive the conclusion of Thm.\ref{thm1}.
By Prop.\ref{Prop1} above, our axiom system \ax{SpecRelDyn} implies
neither \ax{ConsMass} nor \ax{ConsMoment}. By Thm.\ref{thm2},
\ax{ConsMass} and \ax{ConsMoment} together imply the key axiom
\ax{AxCenter} of \ax{SpecRelDyn}. So Thm.\ref{thm1} is stronger than
the corresponding result in the literature since it requires fewer
assumptions.

Thm.\ref{thm2} also states that the conservation axioms can be
replaced by the natural, purely geometrical symmetry postulate
\ax{AxCenter^+} without loss of predictive or expressive power.
Since the conservation axioms \ax{ConsMass} and \ax{ConsMoment} are
not ``purely geometrical'' and they are less observation-oriented
than \ax{AxCenter^+}, we feel that it may be more convincing to use
\ax{AxCenter} or \ax{AxCenter^+} in an axiom system when we
introduce the basics of relativistic dynamics.

\begin{figure}[ht]
\small
\begin{center}
\psfrag*{m}[r][r]{$m_k(b)$} \psfrag*{b}[b][b]{$b$}
\psfrag*{mv}[t][t]{$m_k(b)\cdot\vv_k(b)$}
\psfrag*{P}[lb][lb]{$P_k(b)$} \psfrag*{b}[b][b]{$b$}
\psfrag*{V}[rb][rb]{$V_k(b)$} \psfrag*{v}[rt][rt]{$\vv_k(b)$}
\psfrag*{1}[rt][rt]{$1$} \psfrag*{k}[rt][rt]{$k$}
\includegraphics[keepaspectratio, width=\textwidth]{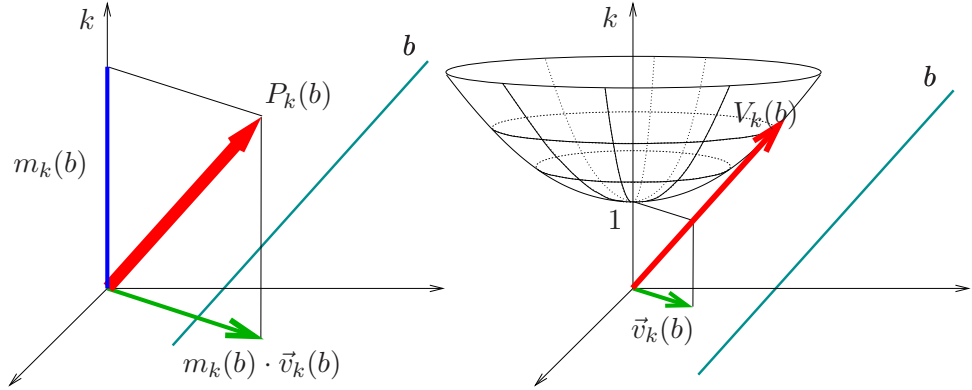}
\caption{\label{fourmoment} Illustration of four-momentum $P_k(b)$ }
\end{center}
\end{figure}

Let $k\in\Ob$ and $b\in\Ib$. The \df{four-momentum} $\Dff{P_k(b)}$
of inertial body $b$ according to observer $k$ is defined to be the
element of $\Q^d$ whose time component and space component are the
relativistic mass and linear-momentum of body $b$ according to
observer $k$, respectively, see Fig.\ref{fourmoment}. That is,
\[
P_k(b)_\tau=m_k(b)\quad \text{ and }\quad
P_k(b)_\sigma=m_k(b)\cdot\vv_k(b).
\]

It is not difficult to see, using Thm.\ref{thm1}, that $P_k(b)$ is
parallel to the world-line of $b$ and its Minkowski-length is
$m_0(b)$.

\begin{description}
\item[\Ax{ConsFourMoment}] Conservation of four-momentum:
\begin{multline*}
\forall k\in\IOb\ \forall b,c,d\in \Ib\quad  [\softcoll_k(b,c:d)\
\Longrightarrow\\
 P_k(b) +P_k(c) =P_k(d)].
\end{multline*}
\end{description}

The following is an immediate corollary of Thm.\ref{thm2}.

\begin{cor}
\label{fourcora} $\ax{AxSelf}\models(\ax{AxCenter^+}\Longleftrightarrow \ax{ConsFourMoment})$.
\end{cor}

Let us return to discussing the merits of using \ax{AxCenter^+} in
place of the more conventional preservation principles. In the
context of Cor.\ref{fourcora}, \ax{ConsFourMoment} has the advantage
that it is computationally direct and simple, while \ax{AxCenter^+}
has the advantage that it is more observational, more geometrical,
and more basic in some intuitive sense.

Let us finally state a theorem about the existence of nontrivial
models of our axiom systems.
\begin{prop}
\label{konzisztencia} \label{rem}
$\ax{SpecRelDyn^+}\cup\{\IOb\neq\emptyset\}$ is consistent.
\end{prop}
\noindent The proof of Prop.\ref{konzisztencia} is in \cite{dyn}.
\medskip

A related work with somewhat different aims is \cite{Reldyn}.

\section{Concluding remarks}
We have introduced a purely geometrical axiom system of special
relativistic dynamics which is strong enough to prove the formula
connecting relativistic and rest masses of bodies. We have also
studied the connection of our key axioms \ax{AxCenter} and
\ax{AxCenter^+} and the usual axioms about the conservation of mass,
momentum and four-momentum. Connections with Einstein's $E=mc^2$
were also discussed. The contents of the present paper represent
only the first steps towards a logical conceptual analysis of
relativistic dynamics or mechanics. A glimpse to Chap.6 (pp.108-130)
``Relativistic particle mechanics" of the textbook
Rindler~\cite{Rin} suggests the topics to be covered in future work
in this line. In a direction orthogonal to this, looking at the
logical issues in \cite{pezsgo} and \cite{logst} suggests questions
and investigations to be carried out into the logical analysis of
relativistic dynamics.

In this paper we began axiomatizing dynamics in special relativity.
This axiomatization of dynamics is extended to the theory of
accelerated observers \ax{AccRel}  in \cite{SzekPhD}. (For the FOL
theory \ax{AccRel} we refer to \cite{Twp}.) In a similar spirit,
these ideas can be naturally extended to the FOL theory \ax{GenRel}
of general relativity (see, e.g., \cite{logst}).

$\ax{AxPh}$ reveals that (in our present axiom systems) we think of
photons as ``possible bodies,'' and the real meaning of \ax{AxPh} is
that ``it is possible for a photon to move from $\pp$ to $\qq$ iff
...''.  The situation is similar with axioms \ax{AxThEx},
\ax{Ax\forall inecoll}. So, a notion of possibility plays a role
here. In the present paper we work in an extensional framework, as
is customary in geometry and in spacetime theory. It would be more
natural to treat this ``possibility phenomenon" in a modal logic
framework, and this is more emphatically so in dynamics. It would be
most interesting to explore the use of a modal logic framework in
our logical analysis of relativity theory. \bigskip

\noindent{\bf Acknowledgements.} Thanks go to Zal\'an Gyenis, Leon
Horsten, Thomas Mueller, Adrian Sfarti and Renata Tordai for helpful
and fruitful discussions, suggestions and remarks.

\bibliography{refs}
\bibliographystyle{plain}

\bigskip\noindent
Alfr\'ed R\'enyi Institute of Mathematics\\
of the Hungarian Academy of Sciences\\
Budapest P.O.Box 127, H-1364 Hungary\\
andreka{@}renyi.hu, madarasz{@}renyi.hu, \\
nemeti{@}renyi.hu, turms{@}renyi.hu.

\end{document}